\documentclass{article}

\usepackage[preprint]{neurips_2024}

\usepackage[utf8]{inputenc} 
\usepackage[T1]{fontenc}    
\usepackage{hyperref}       
\usepackage{url}            
\usepackage{booktabs}       
\usepackage{amsfonts}       
\usepackage{nicefrac}       
\usepackage{microtype}      
\usepackage{xcolor}         

\usepackage{graphicx}
\usepackage{float} 
\usepackage{subfigure}
\usepackage{amsmath}
\usepackage{amssymb}
\usepackage{bm}
\usepackage{algorithm}
\usepackage{algpseudocode}
\usepackage{multirow}

\usepackage{appendix}
\usepackage{colortbl}

\usepackage{natbib}
\setcitestyle{numbers,square}

\usepackage{amsthm}  

\theoremstyle{remark}

\title{A Simple Solution for Homomorphic Evaluation on Large Intervals}

\author{%
\href{https://orcid.org/0000-0003-0378-0607}{\includegraphics[scale=0.06]{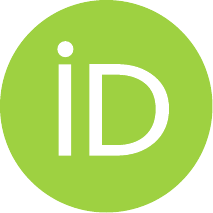}\hspace{1mm}John Chiang } \\
  \texttt{john.chiang.smith@gmail.com} \\
}

\begin{document}
\maketitle

\begin{abstract}
Homomorphic encryption (HE) is a promising technique used for privacy-preserving computation. Since HE schemes only support primitive polynomial operations, homomorphic evaluation of polynomial approximations for non-polynomial functions plays an important role in privacy-preserving machine learning. In this paper, we introduce a simple solution to approximating any functions, which might be overmissed by researchers: just using the neural networks for regressions. By searching decent superparameters, neural networks can achieve near optimal compution depth for a given function with fixed precise, thereby reducing the modulus consumed.

There are three main reasons why we choose neural networks for homomorphic evaluation of polynomial approximations. Firstly, neural networks with polynomial activation functions can be used to approximate whatever funtions needed in an encrypted state. This means that we can compute by one unified process for any polynomail approximation, such as that of Sigmoid or of ReLU. Secondly, by carefully finding an appropriate architecture, neural networks can efficiently evaluate a polynomail using near optimal multicplicative depth, which would consume less modulus and therefore employ less ciphertext refreshing. Finally, as  popular tools, model neural networks have many well-studied techniques that can conveniently  serve our solution.

Experiments showed that our method can be used for approximation of various functions. We exploit our method to the evaluation of the Sigmoid function on large intervals $[-30, +30]$, $[-50, +50]$ and $[-70, +70]$, respectively.

\end{abstract}

\section{Introduction}
Homomorphic encryption (HE) is one of the most popular solutions for privacy-preserving computation, which supports operations over ciphertexts without any decryption process. The privacy-preserving machine learning (ML) applications based on HE, in particular, have gainedd much attention since many ML algorithms are being wildly used for personal information, constantly raising privacy-related issues.

However, it has been a challenging problem to compute a general circuit involving non-polynomial functions in the HE state since HE serves only addition and multiplication operations. The most prevalent solution for this is to replace non-polynomial functions with their polynomial approximations over some ranges. We have to first decide the domain of each real-valued non-polynomial function and then replace the function with its polynomial approximation on the selected interval. In this way, we enable to approximately compute any given circuit in the encrypted doamin, and now one specific prolem left for us is to find an appropriated polynomial approximation for HE.  

At the same time, a computation over encrypted data is different from that in an unencrypted state. That is, the input values to the functions approximated cannot be observed. In addition, the intermediate values during the evaluation cannot be revealed. As a result, there exist only a few clues about the interval for each real-valued function. Therefore, to secure all evaluation results based on HE, we should find polynomail approximations fitting on large intervals.

\subsection{Background}
The approximation error caused by the polynomial approximation affects the quality of computation. When applying a polynomial approximation for the privacy-preserving computation based on HE, we should consider not only the error induced but also the computational cost for the evaluation of the polynomial approximation function. The most desirable approximate evaluation is with a small error and a small computaional cost but in this paper our goal is to minimize the computational cost of approximate evaluation of a real-valued function in an encrypted state, in an tempt to reach a reasonable maximum error.

For the computational cost, the number of multiplications in depth should be taken into account as a top priority since in HE computation the multiplication between two ciphertexts and the boostrapping operation are much heavier than other homomorphic operations. We point out here that a polynomial of high degree does not necessarily need to be evaluated by many number of multiplications. For example, the  polynomial $((x^2 + 1)^2+3)^2$ can be computed with 3 multiplications. Since to approximate a function over a larger interval usually requirs an approximate polynomial of larger degree, we aim to minimize the computational costs during the approximate computation instead of the degree of an approximate polynomial.

To put it all together, the question on desk table is how to evaluate a non-polynomial real-valued function over homomorphically encrypted data with the smallest number of multiplications in depth. However, as we discussed before, HE evaluation requires a sufficiently large domain interval to guarantee the success of the evaluation, which would result in approximation polynomials of suffciciently or even too large degree. Although there have been some general solutions for this question, they introduce too much computational overhead when employing the sufficiently large domain intervals. In this paper, we provide a simple solution to the question. For instance, our solution enables us to evaluate the Sigmoid function at large intervals with reasonalbe modulus costs in encrypted states. With carefully searched parameters, our simple solution to evaluate approximation polynomials requires only $O(\log{}R)$ multiplication depth to approximately evaluate the Sigmoid function on domain interval $[-R, +R]$ with a reasonable maximum error in encrypted states.

More generally, our methodology serves an efficient privacy-preserving evaluation of functions that don't need to converge over the intervals to approximate on. By using our methodology, we can search the desirable polynomial approxmation candidate via the substantial computation resource and evaluate those resulting functions on $[-R, +R]$  with $O(\log R)$ multipliction depth in encrypted states.

\subsection{Related work}

For the privacy-preserving computation, many research works have applied HE to ML algorithms, via appropriate replacement of real-valued functions with their polynomial approximations. Among these HE-based solutions for privacy-preserving ML algorithms,there have been several works that leverage HE for the homomorphic logistic regression~\cite{cheon2018ensemble, kim2018logistic, chen2018logistic, han2019logistic, kim2018secure, IDASH2018bonte, IDASH2018gentry, IDASH2019kim, chiang2022privacy} due to its simple structure.

There also have been many works that utilized HE to other ML algorithms, such as CNN inference~\cite{gilad2016cryptonets, hesamifard2017cryptodl, kim2018matrix, chiang2022volleyrevolver}. In particular, Chiang et al.~\cite{chiang2023privacyCNN, chiang2023privacy3layerNN} developed a novel loss fucntion called Square Likelihood Error that involved Sigmoid function only and presented a solution for neural network training based on mere HE techniques, which with transfer learning can be used to train CNN.

One important related work is the paper of Cheon et al.~\cite{cheon2022homomorphicevaluation}. They develop Domain  Extension Polynomials. This method achieved a remarkable performance in terms of operations and memory. However, the main limitiation of their method is that DEPs might not have a desirable perfemance when it comes to approxmiating functions other than Simgoid-like functions.




\section{Preliminaries}

\subsection{ Homomorphic Encryption }
HE is one special encryption scheme in that it allows operations in encrypted states without the need of decrypting the data nor of requiring access to the  secret key. Various HE schemes have been presented since Gentry~\cite{gentry2009fully} tackled the over three decades problem of fully homomorphic encryption, one of which is the CKKS scheme~\cite{cheon2017homomorphic} widely adopted in the aplication of machine learning algorithms.  CKKS scheme has a strong advantage in application to machine learning algorithms because it encodes real numbers into plaintexts and therefore it has been adopted in many implementations of privacy-preserving machine learning algorithms based on HE. Like other HE libraries, their open-source library,   $\texttt{HEAAN}$, also supports the Single Instruction Multiple Data (aka SIMD) manner~\cite{SmartandVercauteren_SIMD} to encrypt multiple values into a single ciphertext. 

CKKS scheme first encodes the message, a complex number or a list of complex numbers, into the plaintext  $m$ and then encrypts this plaintext into a ciphertext $\texttt{ct}$ with a public key $pk$: $Enc_{pk}(m) = \texttt{ct}$. The corresponding secret key $sk$ is needed for the decryption process: $Dec_{sk}(ct) = m + e,$ where $e$ is a small error. Please refer [to] \cite{cheon2017homomorphic} for the detail of CKKS scheme. Supposing that $Enc_{pk}(m_1) = \texttt{ct}_1$ and $Enc_{pk}(m_2) = \texttt{ct}_2$, the followings are some main operations used in this paper:
\begin{enumerate}
  
  \item \texttt{Add}($\texttt{ct}_1$, $\texttt{ct}_2$): returns a new ciphertext that encrypts the message $m_1 + m_2$.
  
   \item \texttt{cAdd}($\texttt{ct}_1$, $c$): returns a new ciphertext that encrypts the message$m_1 + c$, where $c$ is a complex number.
   
  \item \texttt{Mul}($\texttt{ct}_1$, $\texttt{ct}_2$): returns a new ciphertext that encrypts the message $m_1 \times m_2$.
  
  \item \texttt{cMul}($\texttt{ct}_1$, $c$): returns a new ciphertext that encrypts the message$m_1 \times c$, where $c$ is a complex number.
  
  \item \texttt{iMul}($\texttt{ct}_1$): returns a new ciphertext that encrypts the message$m_1 \times i$, where $i$  is the square root of $-1$.

  \item \texttt{Neg}($\texttt{ct}_1$): returns a new ciphertext that encrypts the message$- m_1$.

\end{enumerate}

The noise in ciphertexts will be increases along with each multiplicaiton and bootstrapping operation should be adopted to refressh the noise level after consuming some multiplicative levels. Moreover, bootstrapping and multiplication operations are much slower than other opearations. As a concequense, the number of multiplications and the multiplicative depth mainly affect the computational costs of circuits over CKKS ciphertexts.

When employing CKKS scheme, we should first based on the dataset size and the security parameter  carefully select the parameters $N$ and $Q$, where $N$ is the dimension of the ring $R$ and $Q$ the initial modulus size.

\subsection{ Polynomial Approximation  } 

There are many known approximation Theories and techniques In mathematics focusing on function fitting or polynomial approximation of functions.

\paragraph{Least Squares Fitting}  Taylor expansion provides a precise approximation on a small range close to the point of interest, but their approximation error could drastically increase outside the small range. On the other hand, the least squares fitting polynomial gives a better approximation on the whole range. A polynomial of degree $d$ among polynomials of degree $\le$ $d$ that has the smallest square sum error is a least squares fitting polynomial. While there are various numerical methods to construct the least squares fitting polynomial, many programming languages have functions to fit a polynomial to a set of data points in the least-square sense, such as the function $polyfit$ in both MatLab and Python. The least squares method is widely adopted in real-world applications, such as in~\cite{kim2018logistic, han2018efficient, chiang2022privacy}.

\paragraph{Minimax Approximation}
The minimax approximation algorithm (aka $L^{}$ approximation or uniform approximation) is a method that try to find  a polynomial approximation of a given continuous function $f(\cdot)$ that minimizes maximum error. For instance, given an interval $[a, b]$ and a positive interger $d$, a minimax polynomial approximation algorithm aims to find a polynomial $p(x)$ of degree at most $d$ to minimize $$\max_{a \le x \le b} |f(x) - p(x)|. $$ The popular Remez algorithm is one of the algorithms to find the minimax polynomial approximation. Minimax approximation ensures a good quality of a polynomial approximation at each point of the approximation interval and hence it is reasonable to be adopted in the polynomial approximation for HE computation. Chen et al.~\cite{chen2018logistic} implement HE-based logistic regression training by adopting the minimax polynomial approximation on $[-5, +5]$. 

\paragraph{Fourier Approximation}
Fourier approximation is a method to represent a function as the sum of simple sine waves, which decomposes periodic functions into a sum of sines and cosines. This technique, introduced by Joseph Fourier in the early 19th century, works well as it provides a better uniform approximation if the function is sufficiently smooth (as is the case with the Sigmoid function). Inspired by the bootstrapping technique of~\cite{cheon2018bootstrapping} via the sine function, we can naturally come up with the idea of applying Fourier series to the approximation of any functions, as mentioned in the  work of~\cite{boura2020chimera,boura2019simulating,boura2018high}.

Cheon et al.~\cite{cheon2018bootstrapping} use a very elegant way via Euler's formula ($e^{ix} = \cos x + i\sin x$) to compute sine function with $\texttt{HEAAN}$ library, can be summarized in Algorithm~\ref{ alg:sine computing function }.

\begin{algorithm}[hp]
    \caption{The Sine Computing Function}
     \begin{algorithmic}[1]
        \Require one ciphertext $\texttt{ct}$ encrypting the target input values $X$; the number $t$  for rescaling $X$ by $2^t$; the number $k$ of iterations for computing Taylor seirives of exp; 
        \Ensure the ciphertext encrypting the approximation result $sin(x)$ 
        
        
        \State Set $\texttt{ct} \gets \texttt{cMul}(\texttt{ct}, 1.0/2^t)$
        \Comment{Initialize the weight vector $\boldsymbol{v} \in \mathbb{R} ^{(1+d)} $}
        \State Set $\texttt{ct} \gets \texttt{iMul}(\texttt{ct})$
        \Comment{Initialize the vector $\boldsymbol{w} \in \mathbb{R} ^{(1+d)} $}
        \State Set $\texttt{nect} \gets \texttt{Neg}(\texttt{ct})$

        \State Set $\texttt{ct1} \gets \texttt{Enc}(1)$
        \State Set $\texttt{ctpow} \gets \texttt{Enc}(1)$
        \For{$i := 1$ to $k$}
        	\State $\texttt{ctpow} \gets \texttt{Mul}(\texttt{ctpow}, \texttt{ct})$ 
			\State $\texttt{ct1} \gets \texttt{Add}(\texttt{ct1}, \texttt{cMul}(\texttt{ctpow}, 1.0/i!))$ 
        \EndFor
        
        \For{$i := 1$ to $t$}
        	\State $\texttt{ct1} \gets \texttt{Mul}(\texttt{ct1}, \texttt{ct1})$ 
        \EndFor

        \State Set $\texttt{ct11} \gets \texttt{Enc}(1)$
        \State Set $\texttt{ctpow} \gets \texttt{Enc}(1)$
        \For{$i := 1$ to $k$}
        	\State $\texttt{ctpow} \gets \texttt{Mul}(\texttt{ctpow}, \texttt{nect})$ 
			\State $\texttt{ct11} \gets \texttt{Add}(\texttt{ct11}, \texttt{cMul}(\texttt{ctpow}, 1.0/i!))$ 
        \EndFor
        
        \For{$i := 1$ to $t$}
        	\State $\texttt{ct11} \gets \texttt{Mul}(\texttt{ct11}, \texttt{ct11})$ 
        \EndFor
        
        \State $\texttt{ctres} \gets \texttt{Add}(\texttt{ct1}, \texttt{Neg}(\texttt{ct11}))$ 
        
        \State $\texttt{ctres} \gets \texttt{cMul}(\texttt{ctres}, -0.5)$
        
        \State $\texttt{ctres} \gets \texttt{iMul}(\texttt{ctres})$
   
        \State \Return $ \texttt{ctres} $
        \end{algorithmic}
       \label{ alg:sine computing function }
\end{algorithm}

On the other hand, Boura et al.~\cite{boura2020chimera} compute absolute function and sign function, which correspond to min/max and comparison respectively, over word-wise encrypted numbers by approximating the functions via Fourier series over a target interval. This method has an advantage on numerical stability compared to general polynomial approximation methods: Since Fourier series is a periodic function, the approximate function does not diverge outside of the interval, while approximate polynomials obtained by polynomial approximation methods do. The homomorphic evaluation of the sign function over wide-wise encrypted inputs is also described in~\cite{bourse2018fast}, which implemented the evaluation phase of discretized neural network based on HE. 

Fourier series is an expansion of a periodic function f(x) in terms of an infinite sum of sines and cosines. If a function $f(x)$ has a finite number of jump discontinuities, which is piecewise smooth~\cite{tolstov2012fourier}, then $f(x)$ has a Fourier series. Such a function is called to be piecewise smooth. For a function $f(x)$ of period $2l$, we can form its Fourier series $f(x) \sim \frac{1}{2}a_0 + \sum_{n=1}^{\infty}(a_n\cos{\frac{n\pi}{l}x} + b_n\sin{\frac{n\pi}{l}x})$, where $a_n = \frac{1}{l}\int_{-l}^lf(x)\cos{\frac{n\pi}{l}x}dx, (n=0,1,2,\ldots)$ and
$b_n = \frac{1}{l}\int_{-l}^lf(x)\sin{\frac{n\pi}{l}x}dx, (n=1,2,\ldots)$. 
The sum of its Fourier series will converge to $f(x)$ at the points of continuity and to the arithmetic mean of the right-hand and left-hand limits at the points of discontinuity. In real-world applications, it is usually a must to replace the infinite series $\sum_{n=1}^{\infty}$ with a finite one $\sum_{n=1}^{N}$ to approximate  the original function, where $N$ is a constant.  Since odd functions have a special feature in Fourier series involving only sine function, we can sometime make the target functions be odd by shifting and resizing. For example, we can shift the Sigmoid function downwards by 0.5 to make a new odd function $F(x)$: $F(x) = Sigmoid(x) - 0.5$.

\paragraph{ High-Degree Polynomial Evaluation }
The calculation of a high-order polynomial of degree $d$ requires $O(\log d)$ to $O(d)$ multiplication operations.
How to compute such polynomials mostly determines the computational costs of the evaluation. For example, Paterson-Stockmeyer algorithm requires $O(\sqrt{d})$ multiplications while Horner's method $O(d)$ times. Therefore, Paterson-Stockmeyer algorithm is commonly being choiced when multiplication is significantly more expensive than addition operation, like the case in an HE system. 

In this paper, we study using neural networks for regression to approxmiate (activation) functions. We show that with fine-tuned parameters neural networks can acheive $O(\log d)$ multiplicative levels.

\subsection{Domain-extension Methodology}
Cheon et al.~\cite{cheon2022homomorphicevaluation} develop a new algorithm to improve the accuracy and efficiency of fitting processes on large intervals. They introduce a special kind of polynomials (DEPs) that can extend the domain interval of target functions while preserving these original functions's feature on its original domain interval. For example, by iterating the domain-extension process with DEPs repeatedly, they can extend the domain of a given function by a factor of $K$ with $O(\log K)$ operations while the feature of the original function is preserved in its original domain interval. In summary, their algorithm has two steps: (1) they restrain the inputs to the given functions to a small range $[-r, +r]$; and (2) they evaluate the target fucntions over the small range $[-r, +r]$ with an approprate polynomial approximation over this small range. Their method uniformly approximating the function on $[-R, +R]$ only exploits $O(\log R)$ operations and $O(1)$ memory, which is more efficient than the previous approach such as the minimax approximation and Paterson-Stockmeyer algorithm. By applying DEPs, their method can efficiently evaluate a function convergeing at infinities in an encrypted state.

While this method~\cite{cheon2022homomorphicevaluation} has remarkable performance in terms of operations and memory, or even  may be perfect solution for the approximation of functions like Sgn, Sigmoid and Tanh, it still remains furthur exploration for functions otherwise. And it might not be suitable for certain functions. Therefore, we present a simple solution for unviserval function approximation just using tradional neural networks, serving as a compliment.

\section{Technical Details}
HE supports only addition and multiplication operations, namely polynomial evaluation. When computing a circuit on homomorphically-encrypted data, we need to replace the non-polynomial functions with their corresponding polynomial approximations over some estimated domain interval. Once the inputs to the target functions escape the approximation interval during the computation, these outlier immediately ruins the evaluation process and contaminates the entire computation. To avoid this, we should select in advance the approximation interval as large as enough for each given non-polynomial functions. Unfortunately, larger approximation intervals requires  higher degree of approximate polynomials, and results in significant computational overhead.

Low-degree polynomials can be efficiently evaluated, but in general, they are useful only on relatively small domain intervals and have unexpected behaviour beyond the small interval. In this work, our goal is to utilize neural networks for regression with polynomial activation functions to approximate and evaluate some given non-polynomial functions on large intervals. If we replaced the common activation functions with their corresponding polynomial approximations over a large enough intervals, the outputs of any such neural networks should be seen as polynomails. With the increase of hidden layers, the degree of these output polynomials will increase exponentially. 

\subsection{ Motivation }
Given sufficient network complexity, neural networks can approximate any continuous function to any desired degree of accuracy. This well-known property of neural networks is universal approximation theorem, which implies that even a simple feedforward neural network with one hidden layer can serve as a universal approximator, provided it has enough neurons. Replacing activation functions satisfying the conditions of the universal approximation theorem with their polynomial approximations over certain range, we can apply neural networks to evaluating any real-valued function involved in an HE system with the same calculation process conssiting of mainly matrix multiplication. Therefore, our simple solution using neural networks can be served as a universial approximator in the encrypted domain.

The universal approximation property underscores the theoretical potential of neural networks to model any function, highlighting their versatility and power. However, this power comes with the challenge of interpretability. While neural networks are powerful approximators, their interpretability, or the ability to understand how they make decisions, remains a significant challenge. Some theoretical connections between neural networks and polynomials have been
noted in the literature~\cite{chiang2023activation, cheng2018polynomial, katznelson1961stone, hornik1989multilayer, poggio1994theory}. Chiang~\cite{chiang2023activation} presented a simple
analytic argument that neural networks model an infinite-dimensional super space and each output is a polynomial of infinite degree but able to be approximated by a finite degree polynomial. In this sense, the outputs of neural networks can roughly be seen as polynomials, exactly what can be evaluated in the homomorphically encrypted state.


\subsection{ Neural Networks }
Neural networks are a subset of machine learning models inspired by the human brain's structure and function. They are designed to recognize patterns, make decisions, and predict outcomes by learning from data. Neural networks are particularly powerful in handling complex tasks such as image and speech recognition, natural language processing, and regression problems.

\paragraph{ Architecture }
The architecture of neural networks consists of three main types of layers: 
\begin{enumerate}
    \item The input layer is the first layer of a neural network, receiving the initial data that the network will process. Each neuron in this layer represents one feature of the input data. For example, in a network designed to process images, each neuron in the input layer might represent the pixel intensity value. 
    
    \item Hidden layers are the intermediary layers between the input and output layers. These layers perform various transformations and computations on the input data. A neural network can have one or multiple hidden layers, depending on its depth. Neurons in hidden layers are connected to neurons in the previous and subsequent layers through weighted connections. Activation functions like ReLU (Rectified Linear Unit), sigmoid, or tanh are applied to the outputs of these neurons to introduce non-linearity, enabling the network to model complex relationships in the data. 
    
    \item The output layer produces the final predictions or classifications based on the input data. The number of neurons in the output layer depends on the specific task. For regression tasks, there is typically one neuron providing a continuous value. For classification tasks, the number of neurons corresponds to the number of classes, often using a softmax activation function to output probabilities for each class.
\end{enumerate}

In addition to the aboves, activation functions is also an important part in neural networks. 


\paragraph{ Training Process }
Training a neural network involves adjusting its weights and biases to minimize the error between predicted and actual values. This process consists of several key parts:

\begin{enumerate}
    \item \texttt{Forward Propagation}: During forward propagation, the input data is passed through the network layer by layer, applying the weights, biases, and activation functions to compute the output.
    
    \item \texttt{Loss Function}: The loss function quantifies the difference between the predicted outputs and the actual target values. Common loss functions include Mean Squared Error (MSE) for regression tasks and Cross-Entropy Loss for classification tasks. For regression tasks, the loss function measures the difference between the predicted values and the actual values. Common loss functions include: Mean Squared Error (MSE), Mean Absolute Error and Huber Loss.
    
    \item \texttt{Backward Propagation}: Backward propagation is used to update the weights and biases based on the computed loss. It involves calculating the gradient of the loss function with respect to each weight and bias, then adjusting them in the direction that reduces the loss. Optimization algorithms like Stochastic Gradient Descent, Adam, or RMSprop are commonly used to perform these updates efficiently.
\end{enumerate}

\begin{figure}[htp]
\centering
\includegraphics[width=5in]{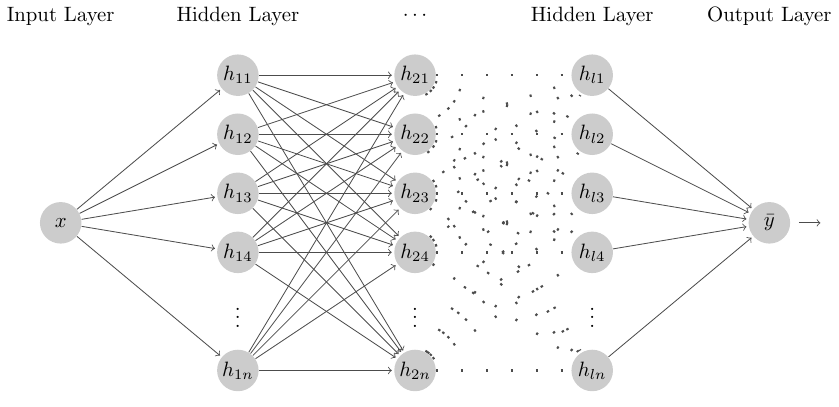} 
\caption{\protect\centering Our unified simple architecture of neural networks for regression tasks}
\label{Neural Networks for Regression}
\end{figure}

\paragraph{ Regression Tasks }
Neural networks are highly effective for regression tasks, where the goal is to predict a continuous output. We adopt a unified simple architecture of neural networks for regression tasks: both the input and output layers have only one node; the hidden layers have $l$ layers each of which all has $n$ nodes, as shown in Figure~\ref{Neural Networks for Regression}.

\subsection{Further Improvements}

\paragraph{ Regularization }
To prevent overfitting, where the network performs well on training data but poorly on unseen data, regularization techniques are applied:

\begin{enumerate}
    \item \texttt{L2 Regularization (Ridge Regression)}: L2 regularization adds a penalty term to the loss function, proportional to the sum of the squared weights, which discourages excessively large weights.
    
    \item \texttt{Dropout}: Dropout randomly sets a fraction of the neurons to zero during training, preventing the network from becoming overly reliant on specific neurons and promoting generalization.
    
    \item \texttt{Early Stopping}: Early stopping monitors the network's performance on a validation dataset during training and stops the training process when performance begins to degrade, indicating potential overfitting.
\end{enumerate}

\paragraph{ Quantization } can be used to optimize the consumed modulus level.
Quantization is the process of mapping input values from a large set (such as real numbers) to output values in a smaller set (such as integers). It is widely used to reduce the size of the neural network model and increase inference speed. There are several types of quantization techniques:

\begin{enumerate}
    \item \texttt{Post-Training Quantization}: This method quantizes a pre-trained full-precision model. It is the simplest approach but may result in some loss of accuracy. It includes techniques like: Static Quantization, Dynamic Quantization and 
Quantization-Aware Training.
    
    \item \texttt{Uniform Quantization}: This method maps input values to uniformly spaced levels. It is simple but may not be optimal for all distributions of weights and activations.
    
    \item \texttt{Non-Uniform Quantization}: This method maps input values to non-uniformly spaced levels, which can be more efficient for certain distributions of weights and activations.
\end{enumerate}

\subsection{ Complexity Analysis }

 Supposing that we are going to approximate some non-polynomial real-valued function $f(x)$ over the interval $[-R, +R]$ with the precision $P$,  the minimum compute unit is $1/P$ and the size of example points to approximate is $2RP$ which can  identify a polynomial of degree at leat $2RP$. Assuming also that we will use a polynomial of degree $d$ as the activation function, then such a neural network with one single hidden layer and enough neruons could represent a polynomial of degree $d$, with two single hidden layers output a polynomial of degree $d^2$, and with $l$ single hidden layers output a polynomial of degree $d^l$. Therefore, the above neural networks with at least $\log ( 2RP/d )$ hidden layers could model the given function $f(x)$ over the interval $[-R, +R]$ with the precision $P$. With fine-tuned parameters, the number of hidden layers can be bound by $O(\log R)$.

\paragraph{ Limitations  }
one limitiation of our solution is that neural networks involve a large number of multiplications and additions, especially for those with many hidden layers. However, we point out here that many of these operations can be computed in parallel and that the multplication depth is desirable due to near optimum complexity.

Anther limitation is that the final polynomial approximataion of our solution is not easy to obtain. We can get the final polynomial by using the Sysbol Package in Python and surely this final polynomial to approximate some given function over a large interval must be a high-degree polynomial with the coefficient to high degrees being too small. Even though we get the polynomial of high degree, how to calculate it is another problem. As a result, there is no pointing to knowing the explicit polynomials.

\paragraph{ Polynomial Function  }
Even though we have a high-order polynomial, how to calculate this polynomial remains a tough task. Paterson-Stockmeyer algorithm or Horner's method can be applied for this problem. However, we can also use neural networks to approximate this high-order polynomial even possibly without any loss of precison. Since the outputs of neural networks with polynomial activition functions are still polynomials, by searching appropriate parameters we can use neural networks to model the target polynomials precisely and perfectly. In this way, we can evaluate the given polynomial with less multiplicative levels, thereby reducing ciphertext modulus and bootstrapping operations.


\section{ Homomorphic Evaluation }
In this section, we explain how neural networks can be employed for privacy-preserving computations based on HE. We mention here that to exploit HE, non-polynomial functions should be in advance replaced by their polynomial approximations, and that the approximation interval should be large enough to contain all input values during the evaluation. Since polynomial approximation on the large intervals, in general, introduces a high degree and significant computational overheads, neural networks may provide a more efficient solution to manage a large domain interval for any target function.

The main calculation for traditional neural networks is matrix multicplication and fortunately there have been many works~\cite{kim2018matrix, chiang2022volleyrevolver} studying efficient matrix multiplciation in the encrypted domain. Given one ciphertext encrypting a list of real numbers, we can just treat it as if it encrypted one row vector. 
The whole pipeline of homomorphic evaluation for some target function $f(x)$ consists of the following  steps:

\indent $\textit{\texttt{ Step 0.}}$ We sample the function $f(x)$ over the approximation intervals, preparing the training dataset.

\indent $\textit{\texttt{ Step 1.}}$ We build the neural network model with polynomial activation functions using some framework such as Tensorflow or PyTorch. 

\indent $\textit{\texttt{ Step 2.}}$ We fit the neural network model into the training dataset with regression loss function like MSE. We tune the parameters for the neural network model and search for a model with decent performance.  

\indent $\textit{\texttt{ Step 3.}}$ After finding an appropriate model with loss function less than a certain error bound, we store the weights of this neural network into a csv file. 

\indent $\textit{\texttt{ Step 4.}}$ For new-coming ciphertext encrypting some real numbers, we compute the neural network model forward  in the encrypted state with the weights abstracted from the above csv file. 

Now, we have the ciphertext encrypting the predictions of function $f(x)$ over the input values encrypted by the orignal ciphertext.

\subsection{Further Improvements}
There is no validate dataset in our solution. However, we could decrease the precision, for example, from $1e-7$ to $1e-1$, and reduce the size of the training dataset. The left unused data point can be therefore used as validate data. Experiments showed that such resulting model shows no worse performance even though we use $1e-1$ to sample our training dataset.

There are also some small tricks to help tune the neural networks. For example, supposing that we can use a $7$-layer neural networks with degree 2 polynimail activation function to approxmate the Simgoid function over the interval $[-30, +30]$, to approximate the Sigmoid function over the interval $[-50, +50]$ we should start test the same neural networks with at least $7$ layers.

After obtaining a decent neural network, we can also use the  Quantization technique to optimize the neural network's weights, such as reducing float weights to integer ones. We admit, however, this technique doesn't help much in our cases, probably due to small room to further optimise our well-tuned models.

\paragraph{Fourier approximation}
By using the trick in bootstrapping to calculate sine funcion, we can also utilize Fourier approximation to approximate any functions. For example, we change the Sigmoid function to a new odd function $F(x) = Sigmoid-0.5$ and obtain its Fourier seriers: $F(x) = \sum_{n = 1}^{\infty} b_n \cdot \sin(\frac{n\pi}{l}\cdot x)$ where $b_n = \frac{2}{l}\int_{0}^l F(x)\sin{\frac{n\pi}{l}x}dx, (n=1,2,\ldots)$ can be computed by Using Simpson's Rule To Calculate the Integration part.

\begin{figure}[t!]
\centering  
\subfigure[Interval 30]{
\label{fig:subfig01}
\includegraphics[width=0.35\textwidth]{  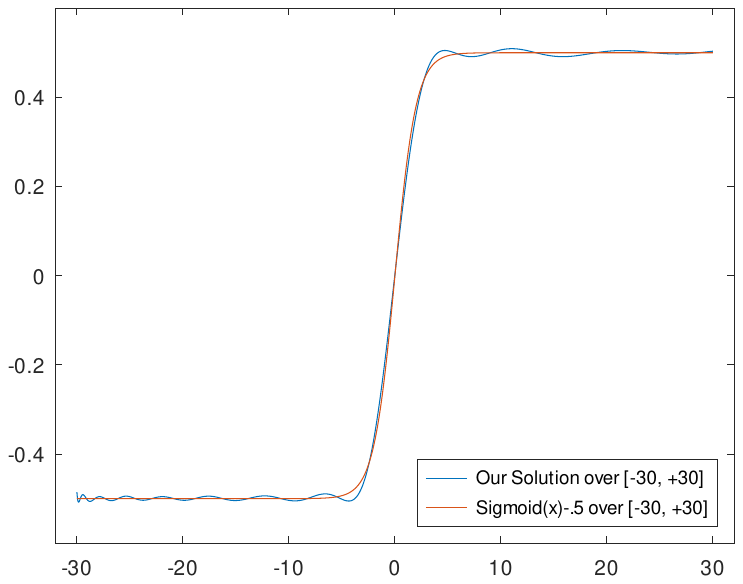  }}
\subfigure[Interval 50]{
\includegraphics[width=0.35\textwidth]{  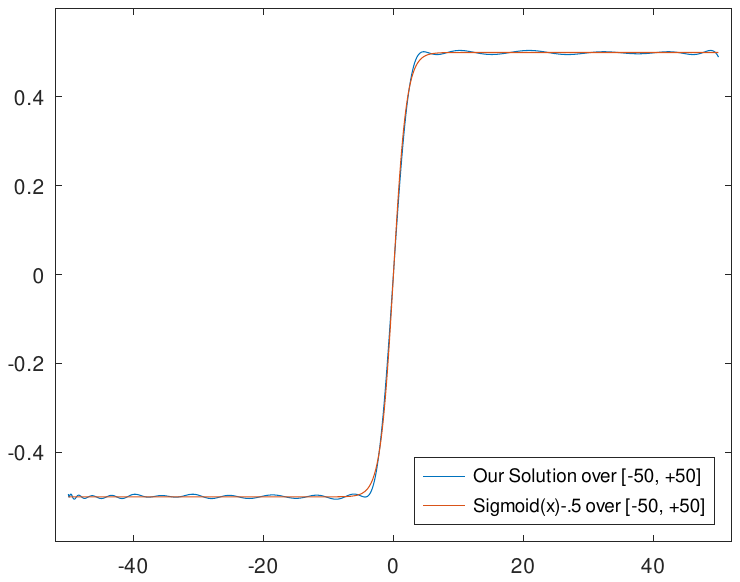  }}
\subfigure[Interval 70]{
\includegraphics[width=0.35\textwidth]{  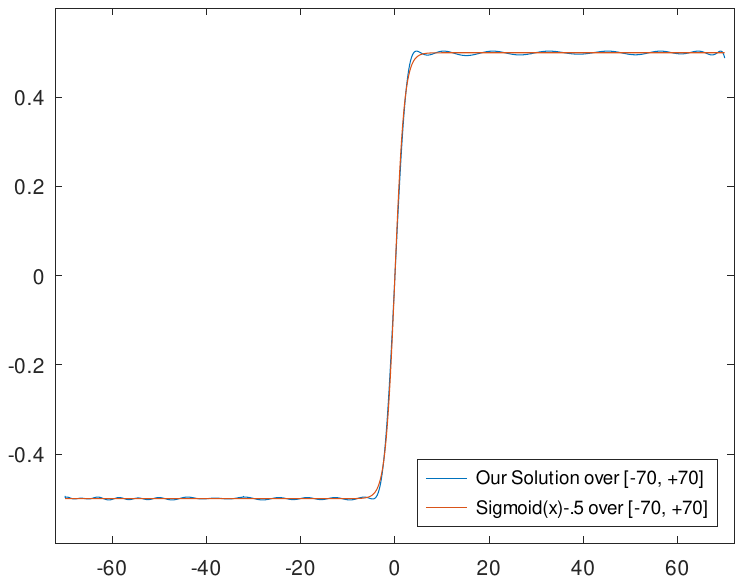  }}
\caption{Our solution over various intervals}
\label{fig:our_solution}
\end{figure}

\section{Experiments}

In this section, we implement our solution using HE with the $\texttt{HEAAN}$ library. The C++ source code is openly accessible at \href{https://github.com/petitioner/HE.HomomorphicEvaluation}{$\texttt{https://github.com/petitioner/HE.HomomorphicEvaluation}$}. 
We use PyTorch to build our neural network model. We approximate the Sigmoid function using neural networks with different number of hidden layers, over the intervals $[-30, +30]$, $[-50, +50]$ and $[-70, +70]$, respectively. The polynomial activation function is initialted as the degree 2 polynomial $p_2 = 1.1110537229 + 0.5 \cdot x + 0.054235537 \cdot x^2  $ from~\cite{chiang2022polynomial} and we set its coeffients to be trainable during the training process. 
The resulting approximation neural networks are then tested with one singal ciphertext encrytping all the real numbers under the given precision ($1e-2$) over the approximation intervals. Figure~\ref{fig:our_solution} shows these results.


\section{Conclusion}

In this paper, we introduced a simple solution for homomorphic evaluation of any function involved in an HE system. 

\bibliography{HomomorphicEvaluation}

\begin{thebibliography}{}

\bibitem[Bonte and Vercauteren, 2018]{IDASH2018bonte}
Bonte, C. and Vercauteren, F. (2018).
\newblock Privacy-preserving logistic regression training.
\newblock {\em BMC medical genomics}, 11(4):86.

\bibitem[Boura et~al., 2018]{boura2018high}
Boura, C., Chillotti, I., Gama, N., Jetchev, D., Peceny, S., and Petric, A.
  (2018).
\newblock High-precision privacy-preserving real-valued function evaluation.
\newblock In {\em Financial Cryptography and Data Security: 22nd International
  Conference, FC 2018, Nieuwpoort, Cura{\c{c}}ao, February 26--March 2, 2018,
  Revised Selected Papers 22}, pages 183--202. Springer.

\bibitem[Boura et~al., 2019]{boura2019simulating}
Boura, C., Gama, N., Georgieva, M., and Jetchev, D. (2019).
\newblock Simulating homomorphic evaluation of deep learning predictions.
\newblock In {\em International Symposium on Cyber Security Cryptography and
  Machine Learning}, pages 212--230. Springer.

\bibitem[Boura et~al., 2020]{boura2020chimera}
Boura, C., Gama, N., Georgieva, M., and Jetchev, D. (2020).
\newblock Chimera: Combining ring-lwe-based fully homomorphic encryption
  schemes.
\newblock {\em Journal of Mathematical Cryptology}, 14(1):316--338.

\bibitem[Bourse et~al., 2018]{bourse2018fast}
Bourse, F., Minelli, M., Minihold, M., and Paillier, P. (2018).
\newblock Fast homomorphic evaluation of deep discretized neural networks.
\newblock In {\em Advances in Cryptology--CRYPTO 2018: 38th Annual
  International Cryptology Conference, Santa Barbara, CA, USA, August 19--23,
  2018, Proceedings, Part III 38}, pages 483--512. Springer.

\bibitem[Chen et~al., 2018]{chen2018logistic}
Chen, H., Gilad-Bachrach, R., Han, K., Huang, Z., Jalali, A., Laine, K., and
  Lauter, K. (2018).
\newblock Logistic regression over encrypted data from fully homomorphic
  encryption.
\newblock {\em BMC medical genomics}, 11:3--12.

\bibitem[Cheng et~al., 2018]{cheng2018polynomial}
Cheng, X., Khomtchouk, B., Matloff, N., and Mohanty, P. (2018).
\newblock Polynomial regression as an alternative to neural nets.
\newblock {\em arXiv preprint arXiv:1806.06850}.

\bibitem[Cheon et~al., 2018a]{cheon2018bootstrapping}
Cheon, J.~H., Han, K., Kim, A., Kim, M., and Song, Y. (2018a).
\newblock Bootstrapping for approximate homomorphic encryption.
\newblock In {\em Annual International Conference on the Theory and
  Applications of Cryptographic Techniques}, pages 360--384. Springer.

\bibitem[Cheon et~al., 2017]{cheon2017homomorphic}
Cheon, J.~H., Kim, A., Kim, M., and Song, Y. (2017).
\newblock Homomorphic encryption for arithmetic of approximate numbers.
\newblock In {\em Advances in Cryptology--ASIACRYPT 2017: 23rd International
  Conference on the Theory and Applications of Cryptology and Information
  Security, Hong Kong, China, December 3-7, 2017, Proceedings, Part I 23},
  pages 409--437. Springer.

\bibitem[Cheon et~al., 2018b]{cheon2018ensemble}
Cheon, J.~H., Kim, D., Kim, Y., and Song, Y. (2018b).
\newblock Ensemble method for privacy-preserving logistic regression based on
  homomorphic encryption.
\newblock {\em IEEE Access}, 6:46938--46948.

\bibitem[Cheon et~al., 2022]{cheon2022homomorphicevaluation}
Cheon, J.~H., Kim, W., and Park, J.~H. (2022).
\newblock Efficient homomorphic evaluation on large intervals.
\newblock {\em IEEE Transactions on Information Forensics and Security},
  17:2553--2568.

\bibitem[Chiang, 2022a]{chiang2022polynomial}
Chiang, J. (2022a).
\newblock On polynomial approximation of activation function.
\newblock {\em arXiv e-prints}, pages arXiv--2202.

\bibitem[Chiang, 2022b]{chiang2022privacy}
Chiang, J. (2022b).
\newblock Privacy-preserving logistic regression training with a faster
  gradient variant.
\newblock {\em arXiv preprint arXiv:2201.10838}.

\bibitem[Chiang, 2022c]{chiang2022volleyrevolver}
Chiang, J. (2022c).
\newblock Volley revolver: A novel matrix-encoding method for
  privacy-preserving neural networks (inference).
\newblock {\em arXiv e-prints}, pages arXiv--2201.

\bibitem[Chiang, 2023a]{chiang2023activation}
Chiang, J. (2023a).
\newblock Activation functions not to active: A plausible theory on
  interpreting neural networks.
\newblock {\em arXiv preprint arXiv:2305.00663}.

\bibitem[Chiang, 2023b]{chiang2023privacy3layerNN}
Chiang, J. (2023b).
\newblock Privacy-preserving 3-layer neural network training using mere
  homomorphic encryption technique.
\newblock {\em arXiv e-prints}, pages arXiv--2308.

\bibitem[Chiang, 2023c]{chiang2023privacyCNN}
Chiang, J. (2023c).
\newblock Privacy-preserving cnn training with transfer learning.
\newblock {\em arXiv e-prints}, pages arXiv--2304.

\bibitem[Crawford et~al., 2018]{IDASH2018gentry}
Crawford, J.~L., Gentry, C., Halevi, S., Platt, D., and Shoup, V. (2018).
\newblock Doing real work with fhe: the case of logistic regression.
\newblock In {\em Proceedings of the 6th Workshop on Encrypted Computing \&
  Applied Homomorphic Cryptography}, pages 1--12.

\bibitem[Gentry, 2009]{gentry2009fully}
Gentry, C. (2009).
\newblock Fully homomorphic encryption using ideal lattices.
\newblock In {\em Proceedings of the forty-first annual ACM symposium on Theory
  of computing}, pages 169--178.

\bibitem[Gilad-Bachrach et~al., 2016]{gilad2016cryptonets}
Gilad-Bachrach, R., Dowlin, N., Laine, K., Lauter, K., Naehrig, M., and
  Wernsing, J. (2016).
\newblock Cryptonets: Applying neural networks to encrypted data with high
  throughput and accuracy.
\newblock In {\em International conference on machine learning}, pages
  201--210. PMLR.

\bibitem[Han et~al., 2019a]{han2019logistic}
Han, K., Hong, S., Cheon, J.~H., and Park, D. (2019a).
\newblock Logistic regression on homomorphic encrypted data at scale.
\newblock In {\em Proceedings of the AAAI conference on artificial
  intelligence}, volume~33, pages 9466--9471.

\bibitem[Han et~al., 2019b]{han2018efficient}
Han, K., Hong, S., Cheon, J.~H., and Park, D. (2019b).
\newblock Logistic regression on homomorphic encrypted data at scale.
\newblock In {\em Proceedings of the AAAI Conference on Artificial
  Intelligence}, volume~33, pages 9466--9471.

\bibitem[Hesamifard et~al., 2017]{hesamifard2017cryptodl}
Hesamifard, E., Takabi, H., and Ghasemi, M. (2017).
\newblock Cryptodl: Deep neural networks over encrypted data.
\newblock {\em arXiv preprint arXiv:1711.05189}.

\bibitem[Hornik et~al., 1989]{hornik1989multilayer}
Hornik, K., Stinchcombe, M., and White, H. (1989).
\newblock Multilayer feedforward networks are universal approximators.
\newblock {\em Neural networks}, 2(5):359--366.

\bibitem[Jiang et~al., 2018]{kim2018matrix}
Jiang, X., Kim, M., Lauter, K., and Song, Y. (2018).
\newblock Secure outsourced matrix computation and application to neural
  networks.
\newblock In {\em Proceedings of the 2018 ACM SIGSAC Conference on Computer and
  Communications Security}, pages 1209--1222.

\bibitem[Katznelson and Rudin, 1961]{katznelson1961stone}
Katznelson, Y. and Rudin, W. (1961).
\newblock The stone-weierstrass property in banach algebras.
\newblock {\em Pacific J. Math}, 11:253--265.

\bibitem[Kim et~al., 2018a]{kim2018logistic}
Kim, A., Song, Y., Kim, M., Lee, K., and Cheon, J.~H. (2018a).
\newblock Logistic regression model training based on the approximate
  homomorphic encryption.
\newblock {\em BMC medical genomics}, 11:23--31.

\bibitem[Kim et~al., 2019]{IDASH2019kim}
Kim, M., Song, Y., Li, B., and Micciancio, D. (2019).
\newblock Semi-parallel logistic regression for gwas on encrypted data.
\newblock {\em IACR Cryptology ePrint Archive}, 2019:294.

\bibitem[Kim et~al., 2018b]{kim2018secure}
Kim, M., Song, Y., Wang, S., Xia, Y., Jiang, X., et~al. (2018b).
\newblock Secure logistic regression based on homomorphic encryption: Design
  and evaluation.
\newblock {\em JMIR medical informatics}, 6(2):e8805.

\bibitem[Poggio and Girosi, 1994]{poggio1994theory}
Poggio, T. and Girosi, F. (1994).
\newblock {\em A theory of networks for approximation and learning}.
\newblock Massachusetts Institute of Technology Artificial Intelligence
  Laboratory.

\bibitem[Smart and Vercauteren, 2011]{SmartandVercauteren_SIMD}
Smart, N. and Vercauteren, F. (2011).
\newblock Fully homomorphic simd operations.
\newblock Cryptology ePrint Archive, Report 2011/133.
\newblock \url{https://ia.cr/2011/133}.

\bibitem[Tolstov, 2012]{tolstov2012fourier}
Tolstov, G.~P. (2012).
\newblock {\em Fourier series}.
\newblock Courier Corporation.

\end{thebibliography}
\bibliographystyle{apalike}

\end{document}